\documentclass[12pt, a4paper]{article}
\usepackage[utf8]{inputenc}

\usepackage{times}           
\usepackage[scaled=0.92]{helvet}  

\usepackage[top=2.5cm, bottom=3.5cm, 
            left=2.5cm, right=2.5cm,
            headheight=17pt]{geometry}
\newcommand{\hangpara}[1]{%
    \par\hangindent=1.2em\hangafter=1\noindent#1\par}
    
\usepackage{parskip}                   

\usepackage{amsmath}  
\usepackage{textcomp} 
\usepackage{graphicx}
\usepackage{booktabs} 
\usepackage[font=small, labelfont=bf]{caption}

\usepackage[super,numbers,sort&compress]{natbib}
\usepackage[hypertexnames=false,           
            pdfencoding=auto]{hyperref}    
\makeatletter
\renewcommand\@biblabel[1]{\textsuperscript{#1}}  
\patchcmd{\thebibliography}
  {\section*{\refname}}
  {\section*{\refname}\setlength{\itemsep}{0pt} \setlength{\labelwidth}{0em} \setlength{\leftmargin}{2em}}{}{}
\makeatother

\makeatletter
\renewcommand{\@maketitle}{%
  \newpage
  \null
  \vspace*{-5pt} 
  \begin{center}
    {\fontsize{20}{24}\bfseries\selectfont \@title \par} 
    \vspace{16pt} 
    {\fontsize{15}{16}\selectfont \@author \par} 
    \vspace{12pt} 
    {\fontsize{15}{16}\selectfont \@date}
  \end{center}
  \vspace{12pt} 
}
\makeatother

\begin{document}
\setlength{\parindent}{2em} 
\setlength{\parskip}{0\baselineskip} 
\title{\textbf{The Onset of Metastable Turbulence in Pipe Flow}}
\author{Jiashun Guan\textsuperscript{} ~ and ~ Jianjun Tao\textsuperscript{*}}
\date{March 31, 2025} 
\maketitle

\hangpara{\noindent \textsuperscript{} ~ Department of Mechanics and Engineering Science, College of Engineering, Peking University, Beijing, 100871, China}

\hangpara{\noindent \textsuperscript{*} ~ To whom correspondence should be addressed. Email: jjtao@pku.edu.cn}
\vspace{16ex}

\textbf{
The onset of turbulence in pipe flow has been a fundamental challenge in physics, applied mathematics, and engineering for over 140 years\cite{Reynolds1883,Mullin2011,Eckhardt2007,Wu2023,Avila2023}. To date, the precursor of this laminar-turbulent transition is recognized as transient turbulent spots or puffs\cite{Wygnanski1973,Peixinho2006,Hof2006,Willis2007,Eckhardt2008,Hof2008,Avila2011}, but their defining characteristics --- longevity, abrupt relaminarization, and super-exponential lifetime scaling --- have been lack of first-principles explanations. By combining extensive computer simulations, theory, and verifications with experimental data, we identify distinct puff relaminarizations separated by a critical Reynolds number, which are defined by a noisy saddle-node bifurcation derived from the Navier-Stokes equations. Below the critical number, the mean lifetime of puff follows a square-root scaling law, representing an intrinsically deterministic decay dominated by the critical slowing down. Above the critical value, the bifurcation’s node branch creates a potential well stabilizing the turbulence, while the saddle branch mediates stochastic barrier-crossing events that drive memoryless decay \cite{Eckhardt2007,Faisst2004,Schneider2008,Avila2010} --- a hallmark of metastable states \cite{Avila2023,Barkley2016}. Accordingly, the mean lifetimes are solved theoretically and can be fitted super-exponentially\cite{Hof2008,Avila2011}. By quantifying the deterministic and stochastic components in the kinetic energy equation, the lifetime statistics of puffs are analyzed in a unified framework across low-to-moderate Reynolds number regimes, uncovering the mechanisms governing the transition to metastable turbulence in pipe flows.
}

\clearpage

\setlength{\parskip}{1\baselineskip} %

\begin{figure}[t]
    \linespread{1.} 
    \centering
    \includegraphics[width=\linewidth]{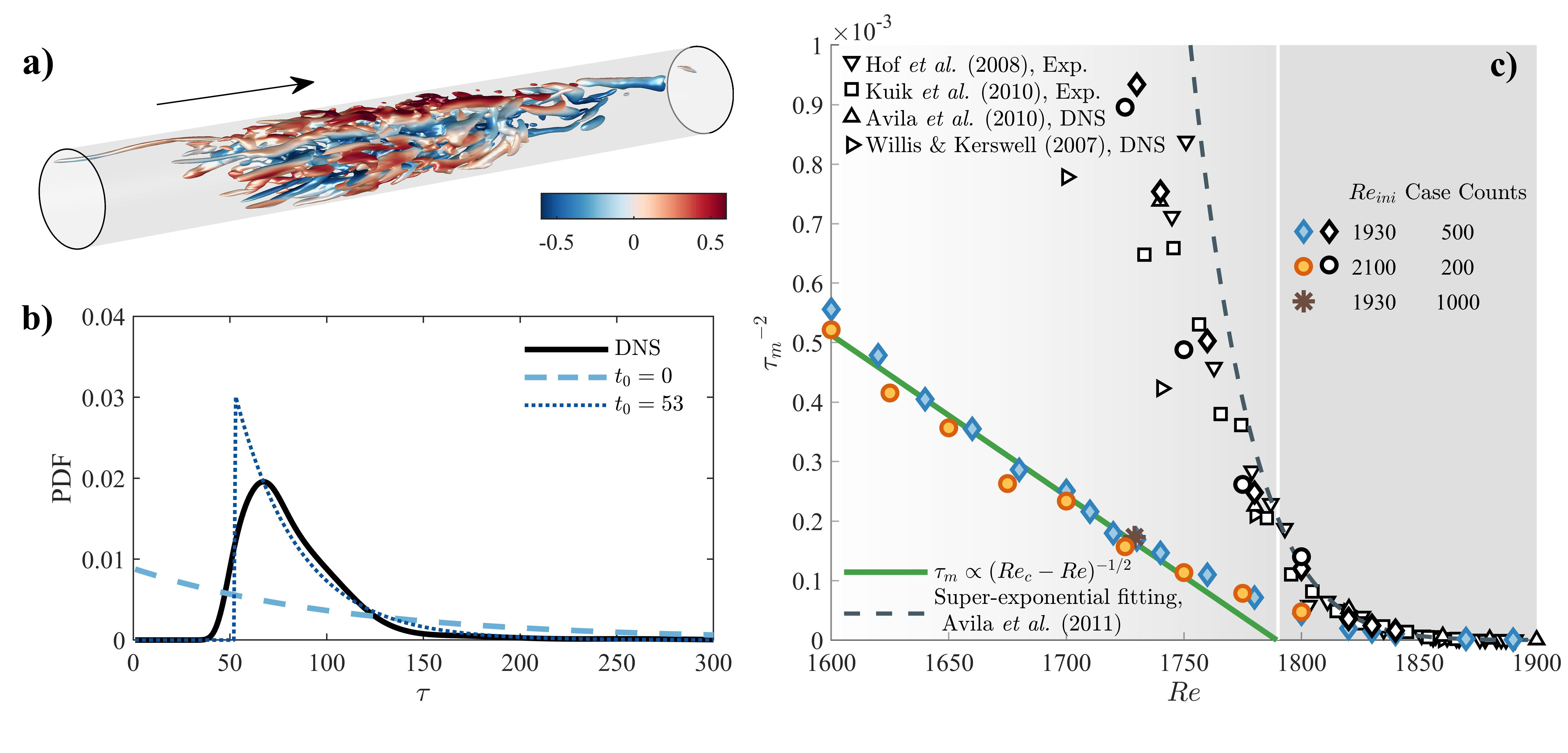}
    \caption{ \textbf{ Lifetime statistics of puff. } \textbf{(a)} A snapshot of turbulent puff at $Re=1940$ with iso-surface of vortex criterion $\Lambda_2 = -0.035$, colored with disturbance streamwise velocity. Flow direction indicated (arrow). \textbf{(b)} Lifetime probability density function (PDF) from 300 DNS cases at $Re=1730$, initialized with developed puffs obtained at $Re_{ini} =1729$. The relaminarization criterion is set as the volume integral of disturbance kinetic energy $E_k<E_{kd}=0.05~(\rho U^2D^3)$, below which $E_k$ decreases monotonically in the explored $Re$ regime (Methods). Exponential fits with $t_0$ are shown for references, and $t_0 = 53$ corresponds to the optimal exponential fit. \textbf{(c)} Mean lifetimes $\tau_m$ versus $Re$. DNS are initialized with puffs obtained at different $Re_{ini}$. Hollow symbols denote $\tau_e$ values from exponential fits. As references, $\tau_e$ of the present data are fitted as well at different $Re$ using the interpolated values of the previously used $t_0$\cite{Avila2010}. The green line is fitted with $\tau_m$ data as $Re<1775$ (Methods).}
    \label{fig:f1}
\end{figure}

Since Reynolds’ pioneering experiments in 1883\cite{Reynolds1883}, the emergence of turbulence in pipe flow has served as a paradigm for studying the laminar-turbulent transition in wall-bounded shear flows \cite{Mullin2011,Eckhardt2007,Wu2023,Avila2023,Barkley2016}. This transition is manifested by localized transient turbulent structures called puffs\cite{Wygnanski1973} (Fig. \ref{fig:f1}a), which exhibit stochastic survival dynamics --- either decaying abruptly or splitting to propagate turbulence\cite{Peixinho2006,Hof2006,Hof2008,Avila2011}. Two fundamental features define the behaviors of puffs, i.e., lifetime statistics follow memoryless exponential distributions\cite{Eckhardt2007,Peixinho2006,Willis2007,Faisst2004,Schneider2008,Avila2010,Kuik2010}, consistent with the chaotic saddle dynamics\cite{Faisst2004,Avila2023,Hof2006,Eckhardt2008,Hof2008}; mean lifetimes exhibit super-exponential growth with increasing the Reynolds number (dimensionless flow rate)\cite{Hof2008,Avila2011}, a scaling proposed to be theoretically linked to extreme value statistics\cite{Goldenfeld2010,Nemoto2021}. Phenomenological models are developed to capture successfully macroscopic features including puff splitting, super-exponential lifetime scaling, and puff jamming\cite{Barkley2016,Barkley2015,Shih2016,lemoult2024directed}. Numerical studies reveal that the exact coherent states (ECS) of pipe flows --- nonlinear solutions of the Navier-Stokes equations with imposed spatio-temporal periodicities and symmetries --- resemble puffs and undergo bifurcation cascades generating transient chaos at low Reynolds numbers\cite{Avila2013}. While the ECS lifetimes similarly exhibit exponential distributions, the transient dynamics in ECS simulations is much simpler than that of puffs\cite{Avila2023}. Despite the successful numerical reproduction of key puff statistics observed experimentally, no first-principles derivation from the full Navier-Stokes equations yet explains these statistical properties, and bridging the gap between the statistical phenomena and the fundamental theory remains crucial for understanding turbulence onset in shear flows.

To statistically characterize puff lifetimes as $Re\le 1980$, we conducted over $14,000$ direct numerical simulations (DNS), with $100\sim 1,000$ cases per Reynolds number (Methods). The probability for flow to retain turbulence beyond time $t$ or the survival probability $S(t)$ is reported to fit an exponential distribution \cite{Eckhardt2007,Peixinho2006,Willis2007,Faisst2004} as $exp[-(t-t_0)/\tau_e(Re)]$, where $\tau_e$ is the characteristic lifetime, $t_0$ denotes the initial adjustment period, and the Reynolds number $Re=UD/\nu$ ($U$: bulk velocity, $D$: pipe diameter, $\nu$: fluid kinematic viscosity). Parameters $U$ and $D$ are set as the characteristic velocity and length scales, respectively. While previously treated as a statistical constant, reported $t_0$ values vary substantially ($50\sim120$) across studies\cite{Hof2006,Willis2007,Hof2008,Avila2011}, e.g., increasing from $72$ at $Re = 1720$ to $95$ at $Re = 1820$ \cite{Avila2010}. We perform DNS at $Re = 1730$ using snapshots of developed puffs (at least 70 $D/U$ away from the initial time and the final relaminarization) obtained at $Re_{ini} = 1729$ as initial fields. Given the minimal Reynolds number relative difference ($\Delta Re/Re < 0.06\%$) and the developed initial puffs, the adjustment period $t_0$ should be statistically negligible. Therefore, the lifetimes of puffs $\tau$ are measured from the initialization to the relaminarization. Figure \ref{fig:f1}b reveals critical limitations of exponential fits at low Reynolds numbers: Both $t_0 = 0$ (dashed curve) and $t_0 = 53$ (dotted curve) fail to capture the main features of  DNS-derived probability density function (PDF) of lifetime, e.g., the smooth ascent near $\tau=40$ and the round peak. The vanishing PDF during the initial phase ($t<40$) mainly reflects not the transient adjustments but an intrinsic inviscid mechanism --- Kelvin's theorem --- preserving the vortical structures in the initial turbulent puffs from decaying. Consequently, we adopt the arithmetic mean lifetimes $\tau_m$ of $\tau$, rather than the exponentially fitted values $\tau_e$, as the characteristic mean lifetime of puff at low and moderate Reynolds numbers. 

In Fig.~\ref{fig:f1}c, a striking divergence between the mean lifetime $\tau_m$ and the exponential-fit $\tau_e$ can be found at low $Re$, and $\tau_m$ exhibits a scaling law ${\tau_m}^{-2}\propto Re_c - Re$ or $\tau_m\propto (Re_c - Re)^{-1/2}$, where $Re_c \approx 1790$ (Methods). The $-1/2$ scaling has not been reported for puffs before, but is found for localized wave packets, the transitional structure in two-dimensional channel flows\cite{Zhang2022}. When $Re$ is larger than $1830$, $\tau_m$ and $\tau_e$ almost collapse with each other. Despite the initial perturbations are puffs obtained at different Reynolds numbers, owning different mean disturbance kinetic energies, the resulting mean lifetimes have no significant differences, e.g.,  $\tau_m=63.2$ and $65.4$ at $Re=1700$ with $Re_{ini}=1930$ and $2100$, respectively, converging to the same scaling law (Fig. \ref{fig:f1}c). When $ Re > Re_c$, the scaling law implies self-sustenance ($\tau_m\rightarrow \infty $), yet $\tau_m$ maintains finite values, and the underlying mechanisms are explained as follows.

\begin{figure}
    \linespread{1.} 
    \centering
    \includegraphics[width=\linewidth]{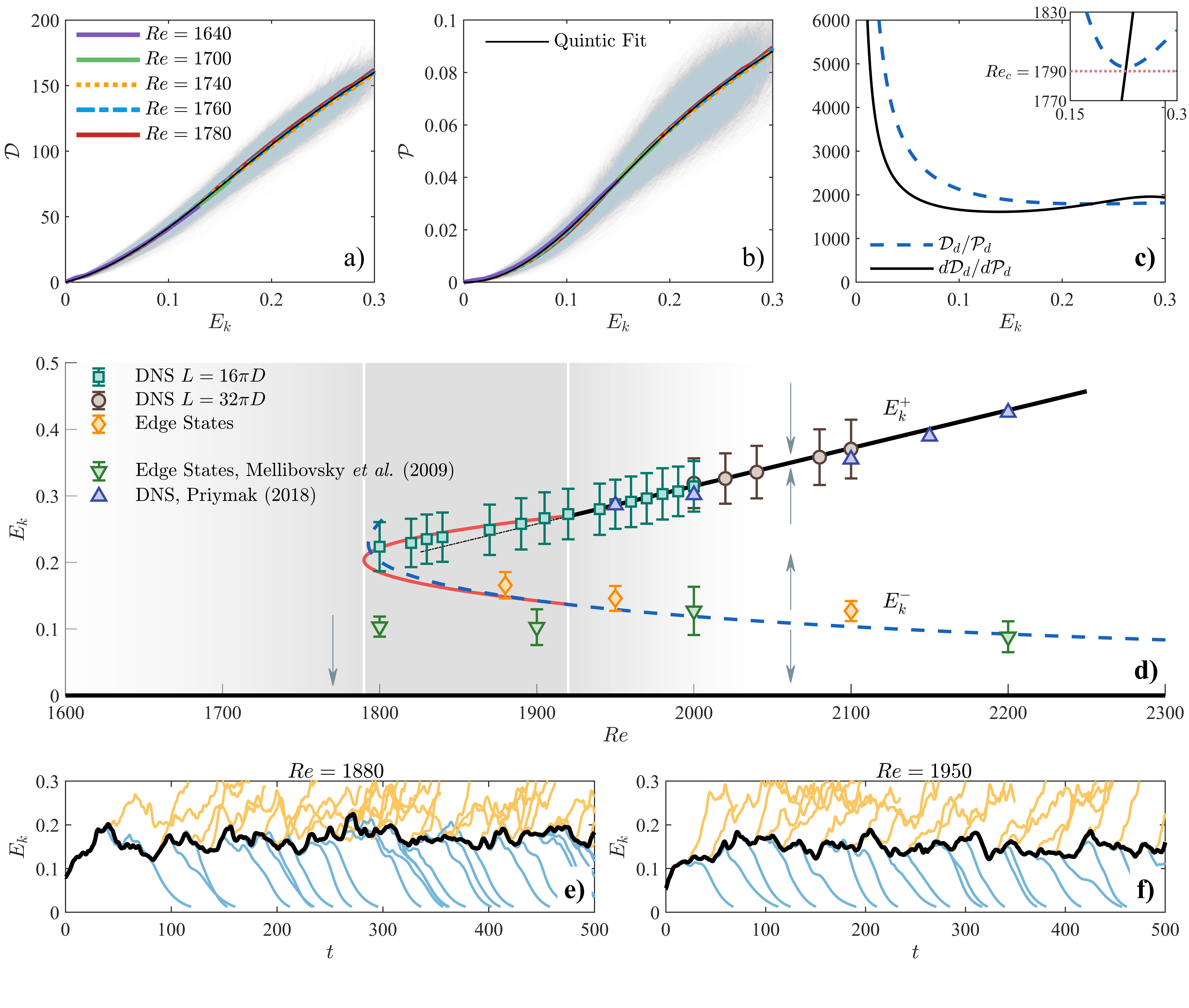}
    \caption{\textbf{Deterministic properties of puff.} \textbf{(a, b)} Phase-space trajectories (gray) of puff in $\mathcal{D}-E_k$ and  $\mathcal{P}-E_k$ coordinates. Ensemble averages at each $E_k$ (Methods) are illustrated as color curves, corresponding to $\mathcal{D}_d$ and $\mathcal{P}_d$ in (a) and (b), respectively, and quantic fit curves (black) accurately capture both trends. \textbf{(c)} Intersection of $\mathcal{D}_d/\mathcal{P}_d$ and $\mathcal{D}_d^\prime/\mathcal{P}_d^\prime$ curves at $\mathcal{D}_d/\mathcal{P}_d$ = $1792.7$. \textbf{(d)} Bifurcation diagram of $E_k$. $E_k^+$ denotes the mean values of $E_k$ excluding the initial and final periods (Methods). The blue dashed line ($E_k^-$) represents the prediction with $Re = \mathcal{D}d/\mathcal{P}d$, where $\mathcal{D}_d$ and $\mathcal{P}_d$ are shown in \textbf{(a)} and \textbf{(b)}, respectively. The red curve follows equation (\ref{eq:3}) within $E_{kt}^- < E_k < E_{kt}^+$ and the arrows indicate the signs of $Re-Rs(E_k)$. The error bars of the present data indicate the standard deviations. \textbf{(e, f)} Edge-tracking via bisection method at $Re = 1880$ and $1950$. Averaged $E_k$ along the thick black trajectories are shown by diamonds in \textbf{(d)}. }
    \label{fig:f2}
\end{figure}

Firstly, the volume integral of disturbance kinetic energy equation derived from the Navier-Stokes equations --- the Reynolds-Orr equation --- is decomposed as 
\begin{equation}
    \frac{dE_k}{dt}=\mathcal{P}-\frac{\mathcal{D}}{Re}=\mathcal{P}_d-\frac{\mathcal{D}_d}{Re}+\varepsilon
    \label{eq:1}
\end{equation}
where $\varepsilon$ represents the contribution of stochastic fluctuations, originated from flow instabilities and strong nonlinear interactions of coherent structures \cite{Manneville2004}, while $\mathcal{P}_d$ and $\mathcal{D}_d/Re$ denote the deterministic components of the energy production term $\mathcal{P}$ and the dissipation term $\mathcal{D}/Re$, respectively. In Fig. \ref{fig:f2}a, phase-space trajectories of $\mathcal{D}$ during the puff evolution reveal that $\mathcal{D}_d$ --- the ensemble average of $\mathcal{D}$ at given $E_k$ --- remains nearly identical across Reynolds numbers, increasing smoothly with $E_k$. For puffs obtained at low Reynolds numbers, $E_k$ seldom reaches high values and hence only the $\mathcal{D}$ and $\mathcal{P}$ with low $E_k$ are considered in the ensemble averages. The deterministic production  $\mathcal{P}_d$ (the ensemble average of $\mathcal{P}$ at given $E_k$) exhibits similar features (Fig. \ref{fig:f2}b). These findings enable us to extend the pattern-preservation approximation originally proposed for the wave packets in channel flows \cite{Zhang2022} to the pipe flows: the general flow pattern is preserved during the structure evolution at different $Re$ and hence the deterministic components $\mathcal{D}_d$ and $\mathcal{P}_d$ depend solely on $E_k$ as shown by the curves in Fig. \ref{fig:f2}a and \ref{fig:f2}b, respectively. 

\begin{figure}[t]
    \linespread{1.} 
    \centering
    \includegraphics[width=\linewidth]{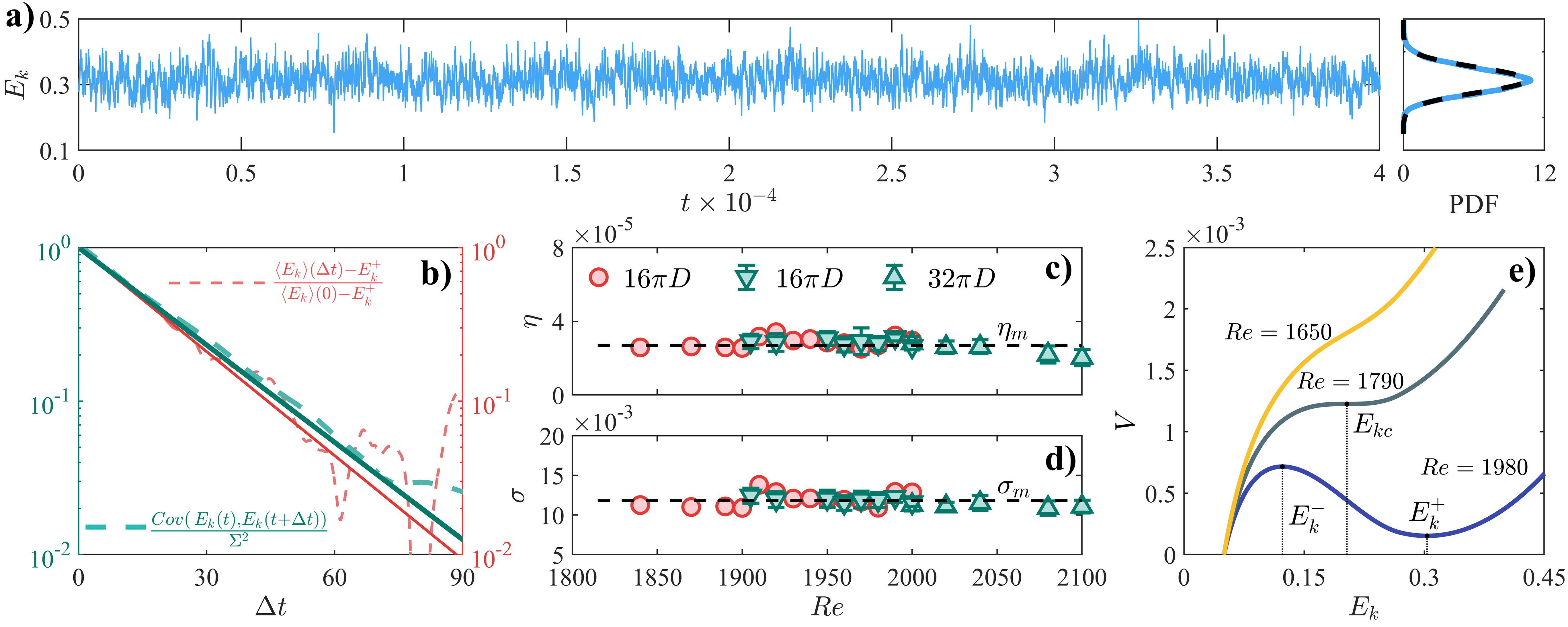}
    \caption{ \textbf{Parameters in the Langevin model.} \textbf{(a)} Temporal evolution of disturbance kinetic energy ($E_k$) at $Re = 2000$, with probability density function (PDF) showing quantitative agreement with Gaussian distribution (dashed curve). \textbf{(b)} Autocovariance normalized by the standard deviation $\Sigma$ (thick green dashed) and ensemble average $\langle E_k \rangle$ (red dashed) derived from the $E_k$ data at $Re = 2000$. The slopes of the fitted solid lines correspond to $-\eta Rs^\prime$ in equation (\ref{eq:5}) (Methods). \textbf{(c)} $\eta$ calculated via autocorrelation decay (green symbols) and ensemble average method (red circles). \textbf{(d)} Stochastic volatility $\sigma =\Sigma \sqrt{2\eta Rs^\prime }$, where $Rs^\prime=dRs/dE_k$ at $E_k=E_k^+ (Re)$. Error bars in \textbf{(c)} and \textbf{(d)} indicate the standard deviations. \textbf{(e)} Potential energy function $V$ versus $E_k$ at different $Re$.}
    \label{fig:f3}
\end{figure}

Secondly, an idealized puff --- deterministic puff, whose integral kinetic energy equation has no fluctuation term $\varepsilon$,  is introduced to analyze the deterministic dynamic behaviors of puff. The counterpart of deterministic puff in two-dimensional channel flows is the localized wave packets\cite{Zhang2022,Zammert2017}. Zero value of $\mathcal{P}_d-\mathcal{D}_d/Re$ represents the zero energy growth rate and hence the steady states of deterministic puffs, whose corresponding Reynolds numbers can be obtained easily as $Rs = \mathcal{D}_d/\mathcal{P}_d$ based on the pattern-preservation approximation. The critical Reynolds number $Re_c$, defined as the minimum $Rs$, emerges as:
\begin{equation}
    Re_c=\frac{\mathcal{D}_d}{\mathcal{P}_d} \text{, where } ~\frac{d(\mathcal{D}_d/\mathcal{P}_d)}{dE_k}=0 ~ \text{ or } ~ \frac{\mathcal{D}_d^\prime}{\mathcal{P}_d^\prime} = \frac{\mathcal{D}_d}{\mathcal{P}_d} 
    \label{eq:2}
\end{equation}
where $^\prime$ denotes $d/dE_k$. As shown in Fig. \ref{fig:f3}c, the $\mathcal{D}_d/\mathcal{P}_d$ and $\mathcal{D}_d^\prime/\mathcal{P}_d^\prime$ curves intersect at $\mathcal{D}_d/\mathcal{P}_d = 1792.7$, agreeing well with $1790$, the $Re_c$ derived from the $-1/2$ scaling shown in Fig. \ref{fig:f1}c, confirming this value as the critical threshold of a bifurcation for sustained deterministic puffs. Consequently, the puff decay at $Re < Re_c$ does not like the behavior of a chaotic saddle \cite{Avila2010,Hof2006,Eckhardt2008,Hof2008,Faisst2004} but corresponds to a deterministic decrease of kinetic energy interfered by stochastic fluctuations. As $Re > Re_c$, the deterministic puff should be self-sustained and the observed transient property of real puffs thus originates from the stochastic contributions.

As $Re > Re_c$, $E_k$ fluctuates around a mean value for a long time (Fig. ~\ref{fig:f3}a), and the mean $E_k$ is estimated as the stable (upper) branch value ($E_k^+$) of the deterministic puff. The fitted relationship follows $E_k^+ =aRe - b = 0.00057Re -0.83$ (black line, Fig. \ref{fig:f2}d) down to $Re > Re_t = 1920$. The dashed curve in Fig. \ref{fig:f2}c in fact predicts the lower branch disturbance kinetic energy $E_k^-$ of the steady deterministic puffs. Between $Re_c$ and $Re_t$, the energy branches are close together and the stochastic fluctuations of $E_k$ make it difficult to accurately obtain $E_k^+$ by time average. It is hard as well to obtain clear front speeds of puffs at $Re$ less than $Re_t$ as reported in the previous experimental and numerical studies \cite{Barkley2015}. Considering that the branch kinetic energies at $Re_t$ ($E_{kt}^+$ and $E_{kt}^-$) are known and $dRs/dE_k=0$ at $Rs = Re_c$, the relation between the branch kinetic energy and $Rs$ can be approximated as a quadratic function as shown by the red curve in Fig. \ref{fig:f2}d. Consequently, equation (\ref{eq:1}) turns to be
\begin{equation}
\begin{aligned}
    \frac{dE_k}{dt}= &  \frac{\mathcal{P}_d}{Re} \left( Re-\frac{\mathcal{D}_d}{\mathcal{P}_d} \right) + \varepsilon\simeq\frac{\mathcal{P}_d}{Re} \left( Re-Rs \right) + \varepsilon \\
     \text{where  } Rs(E_k)= &
    \begin{cases}
    (E_k+b)/a,  & E_k \geq E_{kt}^+ \\
    Re_c+(Re_t-Re_c)\left (\frac{2E_k-E_{kt}^+-E_{kt}^-}{E_{kt}^+-E_{kt}^-}  \right )^2, & E_{kt}^- <E_k< E_{kt}^+ \\
    \mathcal{D}_d/\mathcal{P}_d, & E_k \leq E_{kt}^-
\end{cases}.
\end{aligned}
\label{eq:3}
\end{equation}
According to equation (\ref{eq:3}), the signs of $Re-Rs(E_k)$ represent the growth or decay of a deterministic puff and are shown by arrows in Fig. \ref{fig:f2}d, indicating that $E_k^+$ and $E_k^-$ correspond to the node and saddle branches of the bifurcation, respectively. The predicted $E_k^-$ (blue dashed curve) agrees with the edge-state values obtained through the bisection method \cite{skufca2006edge} (Figs. \ref{fig:f2}e, \ref{fig:f2}f), where the initial perturbations are developed puffs obtained at the same $Re$ but their perturbation amplitudes are scaled to the estimated levels. The previous edge-state data \cite{Mellibovsky2009} are consistent with the predictions as well except at low Reynolds numbers, likely stemming from methodological constraints: the initial disturbances (localized pair of rolls) are not efficient enough to reach the edge states and the simulation period ($250~D/U$) is not sufficiently long at low $Re$ compared to higher-$Re$ regime.

Finally, the contribution of stochastic fluctuations to the energy equation is modeled. The Navier-Stokes equations are differential ones, and then the neighbor transitional velocity fields exhibit temporal correlation at a short time interval, with this correlation decaying progressively as the separation time increases. As demonstrated in Fig. \ref{fig:f3}a, the long-term evolution of kinetic energy conforms to a Gaussian distribution. When the sampling interval $\delta t$ is sufficiently large  (while remaining negligible compared to puff lifetimes) to ensure randomized temporal distribution of $E_k$ values, the integral of the stochastic part in equation (\ref{eq:3}) ($\int_{t-\delta t/2}^{t+\delta t/2}\varepsilon dt$) can be approximated as $\sigma \delta W_t$ --- a Wiener process with a stochastic volatility $\sigma$ \cite{Pope2000}. Under the pattern preservation approximation, we model the deterministic components in equation (\ref{eq:3}) $\int_{t-\delta t/2}^{t+\delta t/2}\frac{\mathcal{P}_d}{Re}(Re-Rs)dt\approx \eta(Re-Rs)\delta t$ to guarantee zero growth rate for the deterministic puffs at both the upper and lower energy branches, where $ E_k=E_k^\pm$ ($Rs=Re$). Consequently, the disturbance kinetic energy equation equation (\ref{eq:3}) is reduced to
\begin{equation}
    \delta E_k=\eta \left( Re-Rs(E_k) \right) \delta t+\sigma \delta W_t=-\frac{dV}{dE_k} \delta t+\sigma \delta W_t,
    \label{eq:4}
\end{equation}
which constitutes a Langevin equation, and the potential energy function is defined as $V=-\int_{E_{kd}}^{E_k}\eta(Re-Rs)  dE_k$ with the relaminarization threshold $E_{kd}=0.05~(\rho U^2D^3)$.

At relatively high Reynolds numbers, $E_k$ fluctuates predominantly around the upper branch $E_k^+$, and $Rs$ may be expanded at $E_k^+ (Re)$ as $Rs=Re+Rs^\prime (E_k -E_k^+)+\cdots$. Neglecting higher-order terms in $(E_k-E_k^+)$, equation (\ref{eq:4}) simplifies to:
\begin{equation}
    \delta E_k\approx -\eta Rs^\prime (E_k-E_k^+ )\delta t+\sigma \delta W_t,   
    \label{eq:5}
\end{equation}
representing an Ornstein-Uhlenbeck type process \cite{Uhlenbeck1930} that captures mean-reverting dynamics \cite{Pope2000}. Both the autocovariance and the ensemble average are calculated based on the $E_k$ time series to estimate $\eta$ and $\sigma$ (Methods) (Figs. \ref{fig:f3}b-\ref{fig:f3}d). As shown in Fig. \ref{fig:f3}c and \ref{fig:f3}d, $\eta$ and $\sigma$ remain approximately constant across the studied $Re$ range, with mean values $2.69\times 10^{-5}$ and $0.0118$, respectively. These values are adopted for the subsequent analyses. Now the potential energy function can be integrated and exhibits (Fig. \ref{fig:f3}e) characteristic barrier-well morphology with a metastable minimum at $E_k^+$ and an activation barrier at $E_k^-$ as $Re>Re_c$. This potential landscape naturally explains that the turbulent puffs will laminarize abruptly sooner or later through a fluctuation-activated barrier crossing mechanism, and identifies $Re_c$ as the critical value for metastable puffs. 

Expanding $Rs$ about the critical state $(Rs,E_k )=(Re_c, E_{kc} )$, where $Rs^\prime=0$, and ignoring higher order terms in $(E_k-E_{kc})$, equation (\ref{eq:4}) reduces to
\begin{equation}
    \delta E_k=\delta (E_k-E_{kc})=\eta \left( (Re-Re_c)-\frac{Rs_c^{\prime\prime}}{2} (E_k-E_{kc} )^2 \right ) \delta t+\sigma \delta W_t
    \label{eq:6}
\end{equation}
where $Rs_c^{\prime \prime}= Rs^{\prime \prime}$ as $E_k=E_{kc}$. This equation governs a noisy saddle-node bifurcation \cite{meunier1988noise,hathcock2021reaction}. It is known that classical saddle-node bifurcations exhibit critical slowing down in the ghost region \cite{Scheffer2009, Scheffer2012anticipating}, yielding the square-root scaling law \cite{Strogatz2000}, which corresponds to $\tau_m \propto (Re_c-Re)^{-1/2}$ as shown by the mean lifetime in Fig. \ref{fig:f1}c. As $Re$ decreases and moves away from $Re_c$, the critical slowing down phenomenon weakens and eventually vanishes when $Re$ drops below 1450 (Methods) --- a value slightly larger than the critical Reynolds number associated with a puff-like relative periodic orbit solution \cite{Avila2013}.

\begin{figure}[t]
    \linespread{1.} 
    \centering
    \includegraphics[width=\linewidth]{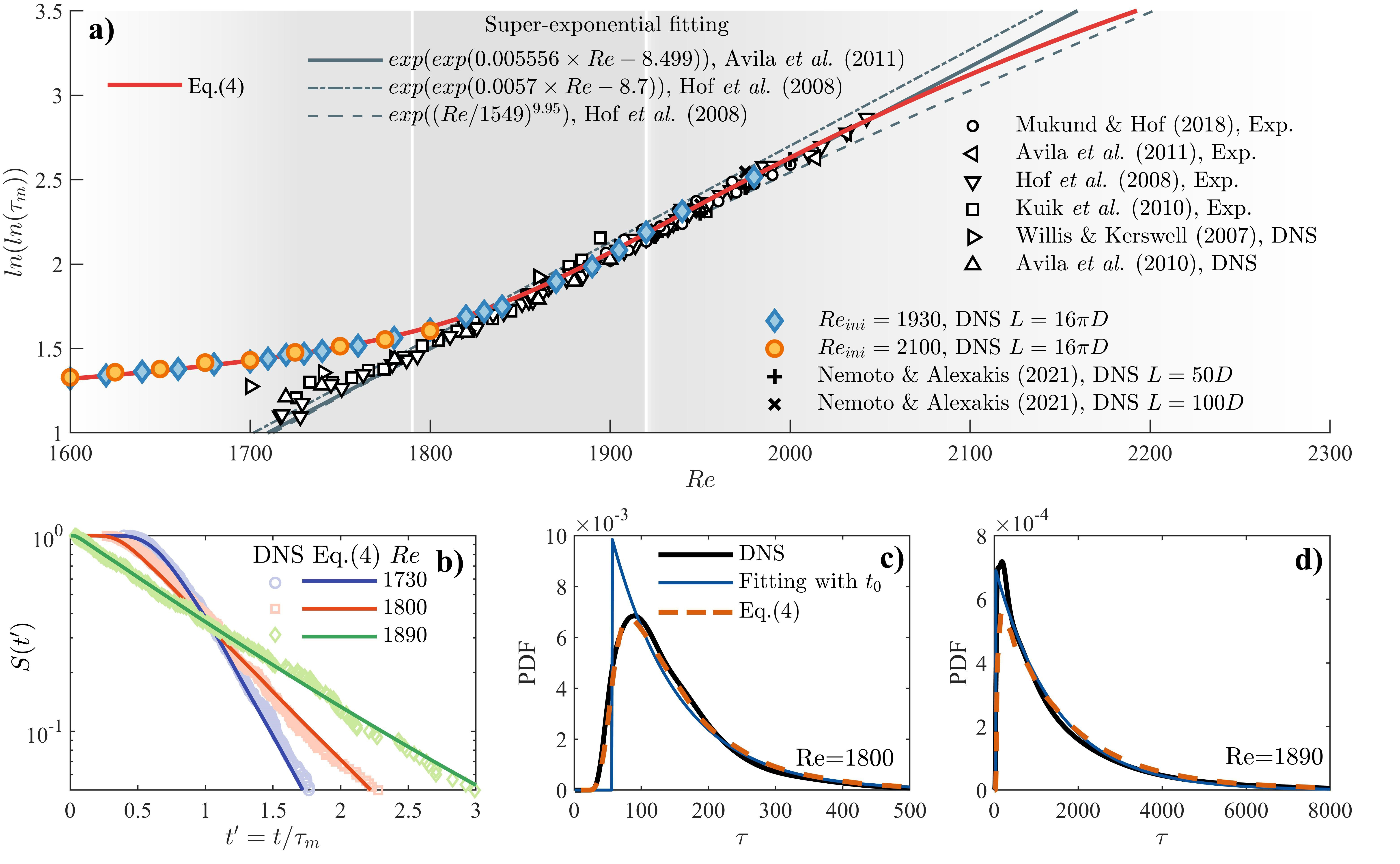}
    \caption{ \textbf{Comparison of lifetime statistics obtained by experiments, simulations, and theory.} \textbf{(a)} Mean lifetime of puff versus $Re$. Hollow symbols denote exponential-fit $\tau_e$ from prior studies\cite{Willis2007,Hof2008,Avila2011,Avila2010,Peixinho2006,Mukund2018}. \textbf{(b)} Survival probabilities from direct numerical simulations (symbols) versus theoretical predictions (curves, equation \ref{eq:4}). \textbf{(c, d)} Lifetime probability density functions at $Re=1800$ and $Re=1890$. Exponential fits optimize $t_0$ to minimize deviations from the DNS data. }
    \label{fig:f4}
\end{figure}

The Langevin framework (equation \ref{eq:4}) enables analytical derivations of puff's lifetime statistics (Methods). Figure \ref{fig:f4}a demonstrates quantitative agreement between the theoretical mean lifetimes ($\tau_m$), predicted with the Langevin model derived \textit{a priori}, and the present DNS data across the moderate and even the low $Re$ regimes. For $Re>1830$, theoretical predictions align with $\tau_e$ obtained by the exponential  fit of experimental/numerical data (Fig. \ref{fig:f4}a), indicating that the stochastic barrier-crossing dynamics dominates the relaminarization at relatively high $Re$. The super-exponential scaling proposed in prior studies \cite{Avila2011} matches the theoretical curve between $Re=1830 \sim 2080$ (Fig. \ref{fig:f4}a). For $Re>2040$, the mean lifetime is extremely high ($\sim 10^8 D/U$) and puff splitting occurs frequently \cite{Avila2011}, precluding reliable decay observations.

In Fig. \ref{fig:f4}b, Survival probabilities $S(t)$ exhibit exponential tails and initial persisting times ($S\approx1$) at low $Re$ (e.g., $Re=1730$), and both characteristics are depicted successfully by the Langevin model. These DNS-derived persisting times, previously attributed to the initial adjustment period $t_0$ and excluded from the lifetime exponential fits\cite{Willis2007,Avila2011,Avila2010}, intrinsically reflect a fact that the initially introduced turbulent structures cannot vanish immediately but persist for at least finite periods, and hence are parts of the lifetimes. For dynamic systems undergoing noisy saddle-node bifurcations, the theory has predicted a continuous deformation of lifetime PDF from Gaussian to exponential distributions across the subcritical-to-supercritical transitions\cite{hathcock2021reaction}. Meanwhile, the corresponding $S(t)$ converts smoothly from a flat-topped curve with an exponential-like tail to a fully exponential distribution, and hence the onset of exponential distribution is not sharp in the control parameter. One may fit the $S(t)$ tail $(t>t_0)$ exponentially even in the subcritical regime so long as $t_0$ is set large enough, e.g., the case with $Re=1730$ in Fig.~\ref{fig:f4}b. Consequently, defining the lower bound for metastable turbulence solely through the exponential tail of $S$ distribution \cite{Avila2010} will systematically underestimate the critical Reynolds number. Lifetime PDFs derived from the Langevin framework show remarkable consistency with DNS data, depicting well the smooth ascent  and the round peak  (Figs. \ref{fig:f4}c). Convergence between the Langevin model and the exponential fit occurs at elevated $Re$, e.g., $Re=1890$, where extended lifetimes minimize the effect of the initial persisting periods on PDF (Fig. \ref{fig:f4}d). 

The dynamical behavior of turbulent puffs is governed by the nonlinear Navier-Stokes equations, with their lifetime statistics captured through the Reynolds-Orr equation, where the energy growth rates emerge from both the deterministic and the stochastic contributions. By applying a pattern-preservation approximation to the deterministic components and modeling the stochastic fluctuations via a Wiener process, the energy equation is transformed into the Langevin equation. With respect to the previously proposed analogies between puffs and dynamic system models \cite{Faisst2004,Barkley2016,Goldenfeld2010,Nemoto2021,Barkley2015,Shih2016,lemoult2024directed}, the Navier-Stokes equations derived Langevin model depicts quantitatively the mean lifetime dependence on the Reynolds number, which can be fitted with the super-exponential scaling in a moderate Reynolds number range.

It is revealed that the subcritical transition in pipe flow originates following a noisy saddle-node bifurcation, characterized by three distinct signatures: (1) a critical Reynolds number $Re_c \approx 1790$ for the metastable puffs, (2) a mean lifetime scaling $\tau_m \propto (Re_c-Re)^{-1/2}$ as $Re < Re_c$, and (3) finite lifetimes governed by the Langevin equation as $Re > Re_c$. This bifurcation framework uncovers fundamentally distinct relaminarization mechanisms of puffs across different $Re$ regimes. In the subcritical regime, the turbulent puffs do not belong to the chaotic saddle: their relaminarizations are intrinsically deterministic, and their mean lifetimes are dominated by the critical slowing down process of the disturbance kinetic energy. In the supercritical regime, a potential energy well and a potential energy barrier are formed in the phase space, corresponding to the upper (node) and lower (saddle) branches of the bifurcation. The stabilizing well maintains puff integrity, explaining prolonged puff persistence at moderate $Re$, while the destabilizing barrier enables memoryless decay through the barrier crossing process driven by stochastic fluctuations, behaving as a chaotic saddle and leading to exponential distributions of lifetime’s survival probability.

As a simplified model of the pipe flow, the subcritical transition in two-dimensional channel flow experiences a saddle-node bifurcation as well \cite{Zhang2022,Zammert2017}, but its integral kinetic energy has no stochastic fluctuations near the critical Reynolds number. Therefore, the global kinetic energy fluctuations are brought by the three-dimensional properties of the pipe flow. Different from the spiral turbulence in three-dimensional Taylor-Couette flows \cite{Coles1965,Feldmann2023} or the oblique turbulent stripes in channel flows \cite{Xiong2015,Tao2018,Tuckerman2020}, which can extend in the azimuthal or spanwise directions, puffs are constrained in both the radial and the azimuthal directions due to the geometry configuration, leading to confined kinetic energies at moderate Reynolds numbers. By verified with experimental and simulation data, the present findings unveil how the intrinsic stochastic fluctuations and the confined global kinetic energies lead to distinct transient properties of puffs, and these encouraging results are expected to advance our understanding of the laminar-turbulent transitions in wall-bounded shear flows.

{\setlength{\parindent}{0pt}\linespread{1.} \fontsize{16}{16}\selectfont \textbf{Acknowledgements}}

This work is supported by the National Natural Science Foundation of China (Grant No. 91752203).

\clearpage
\setlength{\parindent}{0pt}
{\fontsize{20}{16}\selectfont \textbf{Methods}}
\renewcommand{\thetable}{S\arabic{table}} 
\renewcommand{\thefigure}{S\arabic{figure}}  
\renewcommand{\theequation}{S\arabic{equation}}  
\setcounter{figure}{0}
\setcounter{equation}{0}

\section{{ Direct Numerical Simulations of the Navier-Stokes equations}}
Direct numerical simulations (DNS) of the incompressible Navier-Stokes equations are performed for the pipe flows in cylindrical coordinates $(x, r, \theta)$ using a hybrid spectral finite-difference method\cite{willis2017openpipeflow}. The constant flow rate, no-slip boundary conditions on the wall, and streamwise periodic boundary conditions for the disturbance fields (with a pipe length $L$) are imposed, leading to a fixed $Re$ in each numerical case. Spatial discretization was implemented by $n_R$ radial points (located at the roots of a Chebyshev polynomial), $\pm n_M$ azimuthal Fourier modes, and $\pm n_K$ axial Fourier modes. Numerical configurations of DNS cases are summarized in Table \ref{tab:s1}, and the time step is set as $0.0025 ~ (D/U)$. These configurations are well validated in the previous literature \cite{Willis2007,Avila2010,Nemoto2021}.
\begin{table}[hb]
\linespread{1.} 
\centering
\caption{\textbf{Details of DNS.} Type I: $(n_R, n_M, n_K, L)=(40, \pm24, \pm384, 16\pi D)$, type II: $(n_R, n_M, n_K, L)=(64, \pm24, \pm768, 32\pi D)$. The parameter $n$ denotes the total number of runs at the corresponding $Re$ and $T_{max}$ is the maximum simulation time of each case. The group $Re_{L1}$ includes $19$ $Re$s, ranging from $1600$ to $1840$ with an increment of $20$, supplemented by some irregular values: $1400$,  $1500$, $1710$, $1730$, $1830$, and $1870$. The group $Re_{L2}$ includes $15$ $Re$s ranging from $1500$ to $1800$ with an increment of $25$, supplemented with $1400$ and $1450$. Groups $Re_{h1}$ and $Re_{h2}$ are implemented to calculate $\Sigma$, $E_k^+$, and the autocovariance of $E_k$. Specifically, $Re_{h1}$ includes the $Re$ values of $2000$, $2020$, $2040$, $2080$, and $2100$, while $Re_{h2}$ ranges from $1950$ to $2000$ with an increment of $10$. Groups $Re_{ou1}$ and $Re_{ou2}$ are implemented to estimate $\eta$ with the ensemble average method. $Re_{ou1}$ includes cases with $Re=1840$, $1870$, and $1890$, while $Re_{ou2}$ ranges from $1900$ to $2000$ with an increment of $10$. All cases in the left subtable are ended with relaminarization.}
\label{tab:s1}
\begin{minipage}{0.48\textwidth}
\centering
\begin{tabular}{cccc}
\toprule
{\quad $Re$ \quad} & {\quad $Re_{int}$\quad} & {~$n$~} & {\quad Type\quad} \\ 
\midrule
$Re_{L1}$ & \(1930\) & \(500\) & I \\ 
$Re_{L2}$ & \(2100\) & \(200\) & I \\ 
\(1729\) & \(1730\) & \(1000\) & I \\ 
\(1730\) & \(1729\) & \(211\) & I \\ 
\(1890\) & \(1930\) & \(300\) & I \\ 
\(1905\) & \(1930\) & \(100\) & I \\ 
\\
\bottomrule
\end{tabular}
\end{minipage}
\hfill
\begin{minipage}{0.48\textwidth}
\centering
\begin{tabular}{cccc}
\toprule
{~$Re$~} &  {~$Re_{int}$~} &  {\quad $n \times T_{max}$ ~} &  {~Type~ } \\ 
\midrule
\(1920\) & \(1930\) & \(100 \times 8000\) & I \\ 
\(1940\) & \(1930\) & \(150 \times 5000\) & I \\ 
\(1980\) & \(1930\) & \(150 \times 8000\) & I \\ 
$Re_{h1}$ & \(2050\) & \(4 \times 10000\) & II \\ 
$Re_{h2}$ & \(1930\) & \(4 \times 15000\) & I \\ 
$Re_{ou1}$ & \(1990\) & \(100 \times 200\) & I \\ 
$Re_{ou2}$ & \(1840\) & \(100 \times 200\) & I \\ 
\bottomrule
\end{tabular}
\end{minipage}
\end{table}

For each DNS case, the initial field is a single turbulent puff obtained at $ Re_{ini}$ , where the initial puffs are selected to be $10~(D/U)$ from each other and at least $72~(D/U)$ from the beginning and the end of the simulation. The relaminarization criterion is set as the volume integral of disturbance kinetic energy $E_k < E_{kd} =0.05~(\rho U^2D^3)$, below which $E_k$ decreases monotonically as shown below.

\begin{figure}[h]
    \linespread{1.} 
    \centering
    \includegraphics[width=0.5\linewidth]{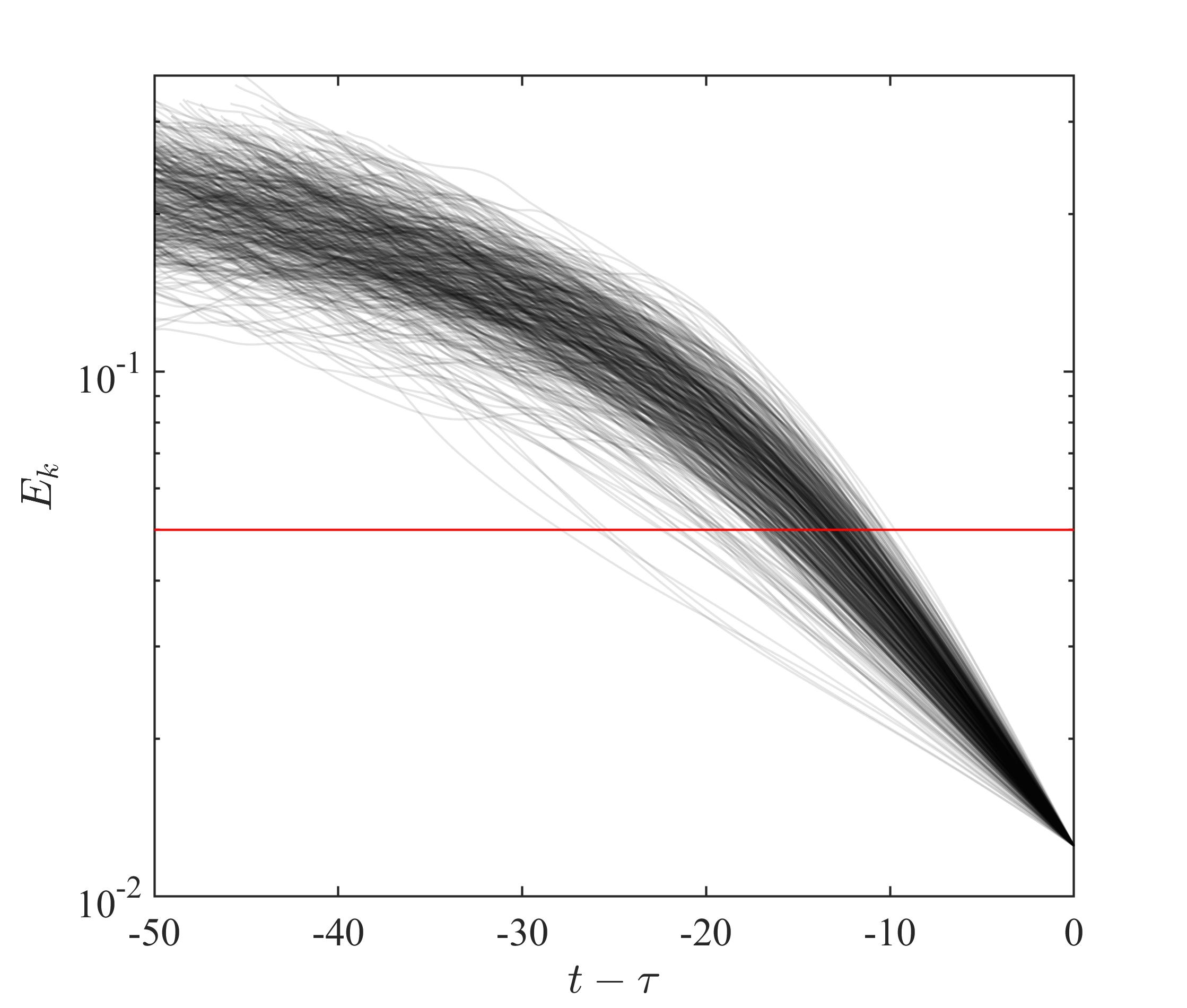}
    \caption{$E_k$ during the relaminarization at $Re=1600$. Totally 500 cases with the initial puffs obtained at $Re_{ini}=1930$. The time $\tau$ corresponds to $E_k=0.0125 ~(\rho U^2D^3)$.}
    \label{fig:s1}
\end{figure}

\section{Kinetic energy equation}
Due to the periodic boundary conditions in the streamwise direction and the nonslip boundary conditions at the cylindrical wall, the integral kinetic energy equation derived from the Navier-Stokes equations, the Reynolds-Orr equation, is decomposed as:
\begin{equation}
\left\{\begin{array}{c}\frac{d E_{k}}{d t}=\mathcal{P}-\frac{1}{Re}\mathcal{D}=\left(\mathcal{P}_d-\frac{1}{Re}\mathcal{D}_d\right)+\varepsilon ,\\ 
E_k = \frac{1}{2}\int_{v} u_i^\prime u_i^\prime dv ,\\
\mathcal{P}=-\int_{v} u_{i}^{\prime} u_{j}^{\prime} \frac{\partial \boldsymbol{U}_{i}}{\partial x_{j}} dv,\\ 
\mathcal{D}=\int_{v} \frac{\partial u_{i}^{\prime}}{\partial x_{j}} \frac{\partial u_{i}^{\prime}}{\partial x_{j}} dv, 
\end{array}\right.
\label{eq:s1}
\end{equation}
where $E_k$, $\mathcal{P}$ and $\mathcal{D}/Re$ are volume-integrated values representing the disturbance kinetic energy, the production term, and the dissipation term, respectively. $\boldsymbol{u}^\prime$ and $\boldsymbol{U}$ denote the disturbance velocity and the base-flow solution, respectively. 

The variables $\mathcal{P}_d$  and  $\mathcal{D}_d$ are the deterministic components of $\mathcal{P}$ and $\mathcal{D}$, respectively. At a given Reynolds number, they are calculated as the ensemble averages of $\mathcal{P}$ and $\mathcal{D}$ within every $E_k$ slice of width $0.01$ in the phase spaces $\mathcal{P}$-$E_k$ and $\mathcal{D}$-$E_k$ (Fig. ~\ref{fig:f2}a, \ref{fig:f2}b), respectively. Because the $\mathcal{P}$ and $\mathcal{D}$ trajectories scatter in the phase spaces, it is validated that the variations of $\mathcal{P}_d$  and  $\mathcal{D}_d$ caused by applying different slice widths ($0.002 \sim 0.02$) are negligible. As shown in Fig. ~\ref{fig:f2}a and \ref{fig:f2}b, the quantic curves fitted by the least squares method agree well with the $\mathcal{P}_d$  and  $\mathcal{D}_d$ curves, and hence are used to calculate $d\mathcal{D}_d/d\mathcal{P}_d =(d\mathcal{D}_d/dE_k)/(d\mathcal{P}_d/dE_k)$ in Fig. ~\ref{fig:f2}c.

The mean values, $E_k^+$ in Fig. \ref{fig:f2}d, are calculated with $E_k$ data $100 ~(D/U)$ away from the initial and the final values (or the relaminarization criterion $E_{kd} =0.05~(\rho U^2D^3))$ in the $E_k$ series, whose total duration spans from $10^3$ to $6\times 10^4 ~ (D/U)$, depending on the Reynolds numbers and case numbers shown in Tab. \ref{tab:s1}.  

\section{Lifetime}
The lifetime of each DNS case is counted from the beginning to the time when $E_k=E_{kd}$ at low Reynolds numbers. At relatively high $Re$, puff lifetimes are long and hence a maximum simulation time $T_{max}$ is applied as $Re \geq 1920$. The mean lifetime is estimated as \cite{Avila2010,Nemoto2021}:
\begin{equation}
    \tau_m=\frac{1}{n_d} \left ( (n-n_d)\times T_{max} +\sum_{i=1}^{n_d} \tau_i \right),
    \label{eq:s2}
\end{equation}
where $\tau_i$, $n$, and $n_d$ represent the lifetime of puff within $T_{max}$, total DNS case number at a given Reynolds number, and the occurrence number of relaminarization, respectively. 

\section{Critical Reynolds Number}
\begin{figure}[t]
    \linespread{1.}
    \centering
    \includegraphics[width=0.6\linewidth]{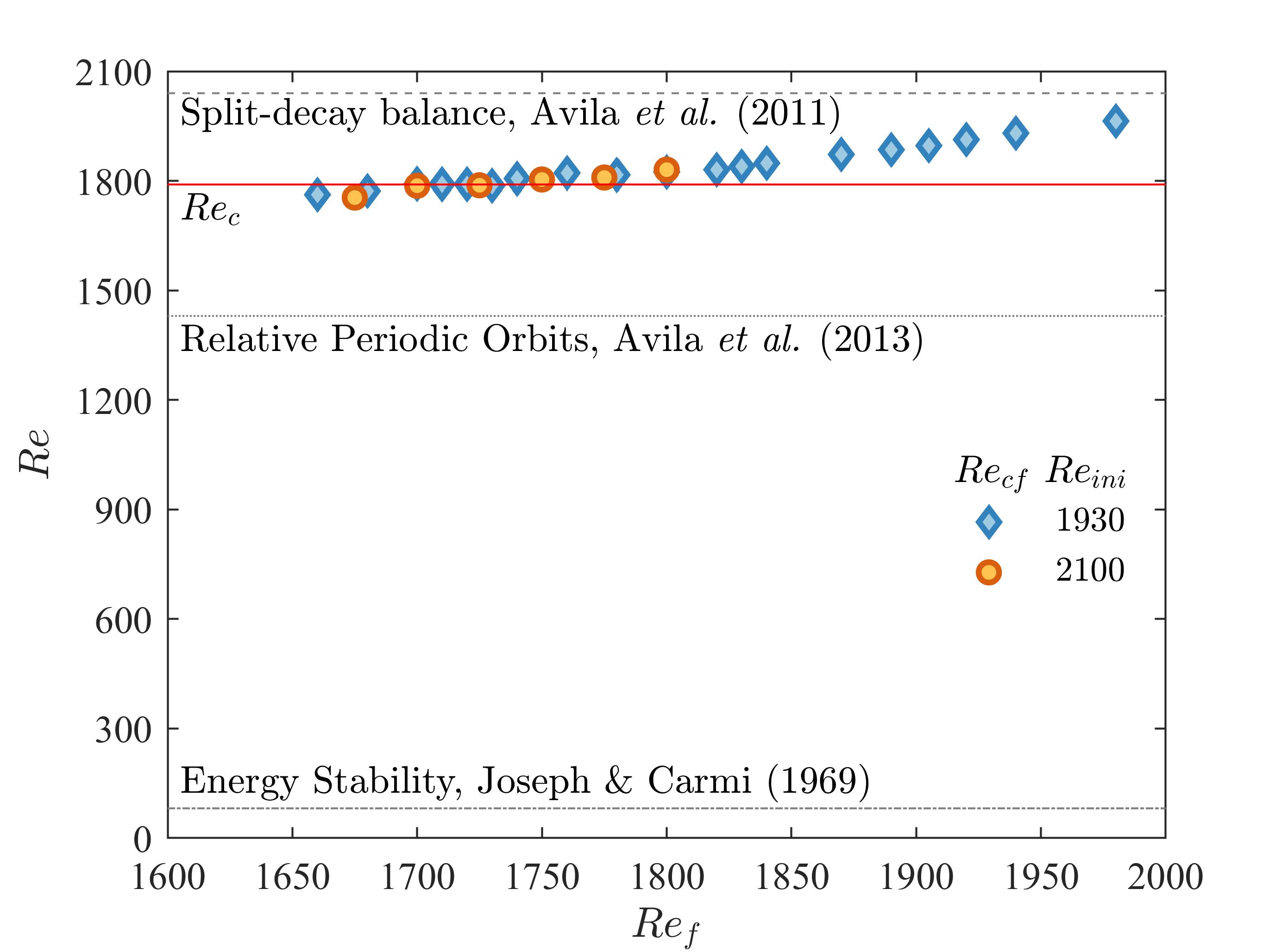}
    \caption{ The critical Reynolds number defined with the square-root scaling. The DNS mean lifetime data (Fig. ~\ref{fig:f1}c) of four adjacent $Re$s ($Re_f$ denotes the maximum $Re$) are linearly fitted as ${\tau_m}^{-2} =A~ Re+B$, and then the critical value $Re_{cf} = -B/A$ are calculated and shown as symbols for different initial puffs obtained at $Re_{ini}$. The $Re$ threshold of energy method \cite{joseph1969stability} and the Reynolds numbers for the relative periodic orbit solution \cite{Avila2013} and the sustained turbulence \cite{Avila2011} are illustrated as well for references.}
    \label{fig:s2}
\end{figure}

At low Reynolds numbers, the initial fields with puffs obtained at $Re_{ini} =2100$ have statistically higher $E_k$ in comparison with those of $Re_{ini}$ =1930, corresponding to slightly longer mean lifetimes (Fig. ~\ref{fig:f1}c). However, these DNS cases share the same square root scaling law: the critical Reynolds numbers $Re_{cf}$ extrapolated from different $Re$ ranges (with the maximum $Re$ denoted as $Re_f$) and $Re_{ini}$ share almost the same value as shown in Fig.~\ref{fig:s2}: $Re_{cf}$ fluctuates slightly around a mean value as $Re_f$ less than 1775, above which $Re_{cf}$ increases monotonically with increasing $Re_f$. The mean value of $Re_{cf}$ as $Re_f<1775$ is 1788.1,  and the critical $Re_c$ obtained by linearly fitting the $\tau_m$ data  as $Re <1775$ (the green line in Fig. 1c) is 1790. Consequently, it is estimated that $Re_c=1790$.

\section{Ornstein–Uhlenbeck process}
Equation (5) in the text,
\begin{equation}
    \delta E_k \approx - \eta Rs^\prime(E_k-E_k^+)\delta t+\sigma \delta W_t,
    \label{eq:s3}
\end{equation}
represents an Ornstein–Uhlenbeck (O–U) process \cite{Uhlenbeck1930}, describing the mean-reverting phenomena with $\eta Rs^\prime$ serving as the speed of mean reversion. According to the properties of O-U process, two methods are used to estimate the values of $\eta$ and $\sigma$. 

\textbf{\underline{Ensemble average method:}} one hundred DNS cases for each $Re$ are implemented in the range of $Re=1840\sim 2000$ with initial puffs obtained at $Re_{ini}=1840$ for $Re\geq 1900$ and $Re_{ini}=1990$ for $Re<1900$. These initial puffs are selected to be $100~(D/U)$ from the initial time and spaced $10~ (D/U)$ apart from each other. Assuming that time series of $E_k$ comply with O-U process, then the ensemble average of $E_k$ at different time $\Delta t$, $\langle E_k\rangle(\Delta t)$, satisfies,
\begin{equation}
\frac{\langle E_k (\Delta t) \rangle-E_k^+}{\langle E_k (0) \rangle-E_k^+}=e^{-\eta Rs^\prime \Delta t}.
\label{eq:s4}
\end{equation}

\textbf{\underline{Autocorrelation method:}} The time series of $E_k$ lasting totally at least $4\times10^4 (D/U)$ (Tab. ~\ref{tab:s1}) are used to calculate its covariance and standard deviation $\Sigma$ to estimate $\eta$:
\begin{equation}
    \frac{Cov\left( E_k(t),E_k(t+\Delta t) \right)}{\Sigma^2}=e^{-\eta Rs^\prime \Delta t}.
\end{equation}

As shown in Fig. ~\ref{fig:f3}b, the data obtained by these two methods are linearly fitted within the time interval $0\sim 20 ~ (D/U)$, and their slopes are shown to be consistent with each other. At relatively high $Re$, the data of $E_k$ used to calculate its mean value $E_k^+$ and standard deviation $\Sigma$ are $100 ~(D/U)$ from the initial time and the time when the relaminarization occurs.

\begin{figure}[h]
	\linespread{1.}
	\centering
	\includegraphics[width=0.95\linewidth]{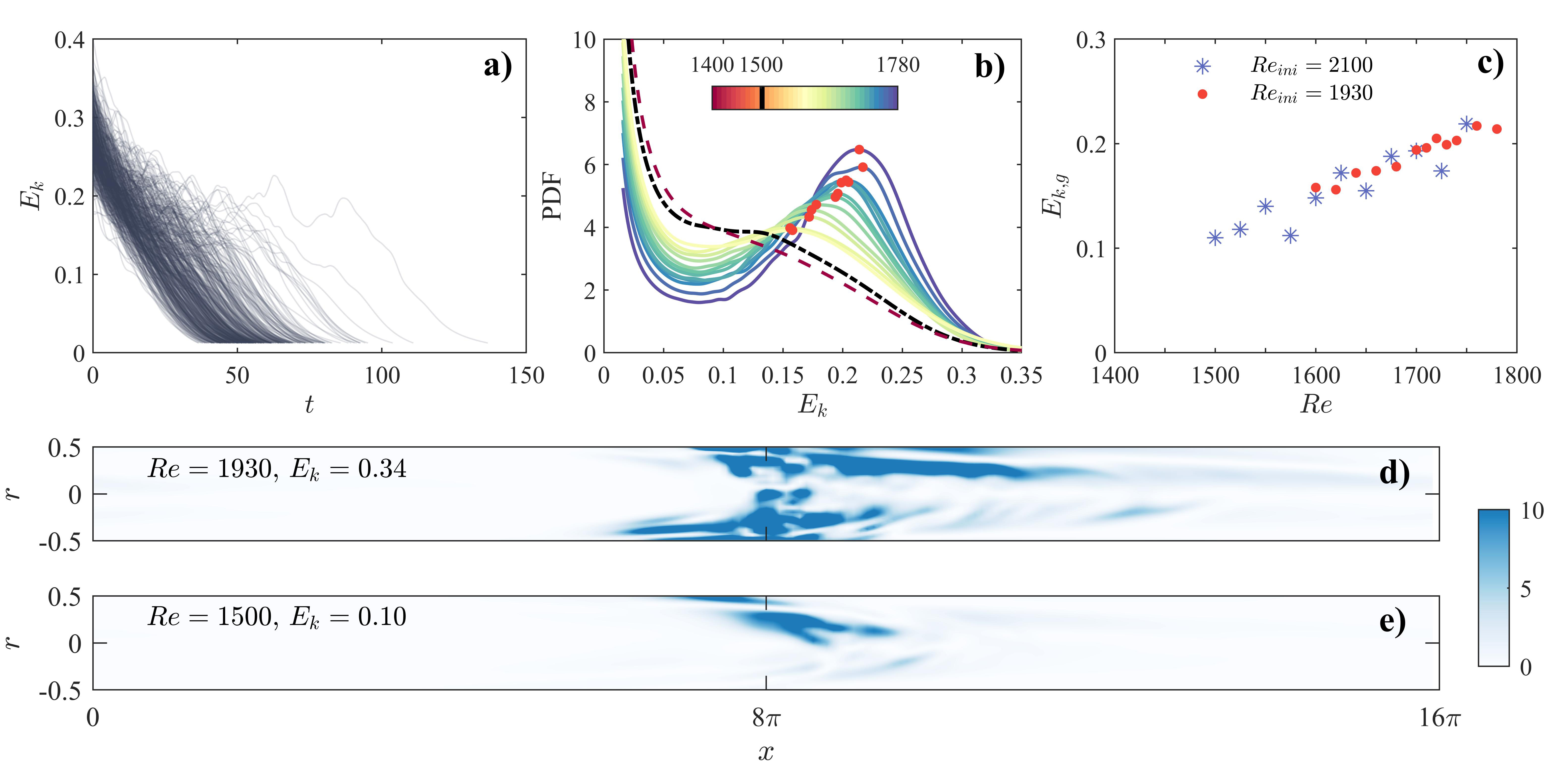}
	\caption{ (a) $E_k$ series at $Re=1600$ and (b) $E_k$'s PDF as a function of $Re$, calculated with 500 cases for each $Re$ initialized with puffs obtained at $Re_{ini}=1930$. The $E_k$ of the maximum PDF, which are labeled by red points and referred as $E_{k,g}$, decrease statically with  reducing $Re$ and disappear at $Re=1450$ as shown in (c), where the data for 200 initial puffs obtained at $Re_{ini}=2100$ are added. Iso-contours of transient disturbance enstrophy at $(Re, E_k)=(1930, 0.34)$ and $(1500, 0.1)$ are shown in (d) and (e), respectively.}
	\label{fig:s3}
\end{figure}

\section{Critical slowing down}

An nonlinear dynamic system approaching a bifurcation point, e.g., $(Re_c, E_{k,c})$ for the pipe flows,  (statistically) exhibits a dramatic increase in the time required to recover from perturbations (the turbulent puff) or return to equilibrium (the laminar state) (Fig.~\ref{fig:s3}a), a phenomenon called critical slowing down \cite{Scheffer2009, Scheffer2012anticipating}. Accordingly, there is a local maximum in the PDF of $E_k$ (Fig.~\ref{fig:s3}b), whose $E_k$ is referred as $E_{k,g}$, corresponding to the statistically minimal decay rate of $E_k$ and reflecting the ghost of saddle-node bifurcation \cite{Strogatz2000}. As shown in Fig.~\ref{fig:s3}b and \ref{fig:s3}c, $E_{k,g}$ is 0.21 at $Re=1780$, which is close to $E_{kc}$, and decreases to about 0.1 at $Re=1500$, where turbulent small scale structures indicated by the disturbance enstrophy seriously diminish, the featured downstream long tail almost disappears, and the flow pattern is different from a complete puff. Below $Re=1450$, no local maxima in $E_k$'s PDF are observed in the test cases and the critical slowing down phenomenon vanishes.

\section{Langevin equation}
The equation (4) in the text,
\begin{equation}
    \delta E_k=\eta \left( Re-Rs(E_k) \right) \delta t+\sigma \delta W_t=-U^\prime \delta t+\sigma \delta W_t,
    \label{eq:s6}
\end{equation}
is the Langevin equation. The mean lifetime of puff corresponds to the mean first passage time of $E_k$  before arriving $E_{kd} =0.05$ (the criterion of relaminarization), and can be analytically derived by utilizing the Kolmogorov backward equation\cite{stratonovich1963topics,hanggi1990reaction}:
\begin{equation}
    \tau_m=\frac{2}{\sigma_m^2} \int_{E_{kd}}^{E_{ki}} e^{\frac{2}{\sigma_m^2} V(\hat{E}_k)} \left(\int_{\hat{E}_k}^{E_{kmax}} e^{-\frac{2}{\sigma_m^2} V(E_k)} dE_k \right )d\hat{E}_k,
    \label{eq:s7}
\end{equation}
where $E_{ki}$ and $E_{kmax}$ represent the initial $E_k$ and the upper boundary of $E_k$ for $V$, respectively, and $\sigma_m$ is the mean value of $\sigma$ shown in Fig. ~\ref{fig:f3}d. As $Re>2250$, puffs evolve into slugs\cite{Barkley2015}, and hence the extrapolated mean value of $E_k$ at $Re=2250$, $E_k^+=0.46$, is set as the $E_{kmax}$. At moderate Reynolds numbers, the statistical properties of puff are insensitive to the initial conditions due to the existence of the potential energy well as shown in Fig. ~\ref{fig:f3}e, and then it is set that $E_{ki}=E_{kmax}$. In addition, the Langevin equation (Equation \ref{eq:4}) is solved numerically by the Euler–Maruyama method\cite{kloeden1992numerical}, employing a time step of $0.02$. Parameters $E_{ki}$ and $E_{kd}$ match those in equation (\ref{eq:s7}). For each $Re$, $10,000$ cases (maximum duration: $3\times10^4 ~ D/U$) are implemented, with lifetime statistics shown in Fig. ~\ref{fig:f4}b-d.

\clearpage
    \bibliography{ref} 
\end{document}